\documentclass[aps,pre,superscriptaddress,twocolumn,showpacs]{revtex4}
\usepackage{amsmath}
\usepackage{graphicx}
\usepackage{amssymb}
\usepackage{bbm, color, soul}

\newcommand*{\tn}[1]{{\textnormal{#1}}}
\newcommand{\ket}[1]{\left|#1\right>}

\usepackage{color}

\begin{document}

\title{The Shannon entropy and avoided crossings in closed and open quantum billiards
}

\author{Kyu-Won Park}
\affiliation{School of Physics and Astronomy, Seoul National University, Seoul 08826, Korea}
\author{Songky Moon}
\affiliation{School of Physics and Astronomy, Seoul National University, Seoul 08826, Korea}
\author{Younghoon Shin}
\affiliation{School of Physics and Astronomy, Seoul National University, Seoul 08826, Korea}
\author{Jinuk Kim}
\affiliation{School of Physics and Astronomy, Seoul National University, Seoul 08826, Korea}
\author{Kabgyun Jeong}
\affiliation{IMDARC, Department of Mathematical Sciences, Seoul National University, Seoul 08826, Korea}
\author{Kyungwon An}
\email{kwan@phya.snu.ac.kr}
\affiliation{School of Physics and Astronomy, Seoul National University, Seoul 08826, Korea}

\date{\today}

\begin{abstract}
The relation between the Shannon entropy and avoided crossings is investigated in dielectric microcavities.
The Shannon entropy of probability density for eigenfunctions in an open elliptic billiard as well as a closed quadrupole billiard increases as the center of avoided crossing is approached. These results are opposite to those of atomic physics for electrons. It is found that the collective Lamb shift of the open quantum system and the symmetry breaking in the closed chaotic quantum system give equivalent effects to the Shannon entropy.
\end{abstract}
\maketitle

\section{Introduction} \label{intro}
The Shannon entropy, first introduced by Shannon in communication theory~\cite{S48}, is a relevant measure of the average amount of information for a random
variable with a specific probability distribution function.
The Shannon entropy is generally equivalent to the von Neumann entropy
for a physical observable with eigenvalues or probability densities.
Not only it is useful in modern information theory but also it has important applications in quantum physics related to uncertainty principle~\cite{HJC17,BM75} and quantum measurement~\cite{Z03,L08} as well as in other areas.
The Shannon entropy has been applied to the identification of putative drug targets in bio-system~\cite{FC00} and descriptor analysis for distinguishing natural products from synthetic molecules~\cite{SG00}, and it was also used to measure topological diversity related to economics~\cite{EM10} and to detect defects in acoustic emission~\cite{ZF15}.
Recently, the Shannon entropy has also been used as an indicator for avoided crossing in atomic systems~\cite{GD033,HC15}.

The avoided crossing is a phenomenon where the two eigenvalues of a Hamiltonian come close and then repel each other as a system parameter is varied. It signifies the presence of an interaction between states in the Hamiltonian. By this reason, the avoided crossing has been a fundamentally important concept~\cite{NW29} from the beginning of quantum mechanics.
It has been extensively studied theoretically as well as experimentally in various physical systems~\cite{CAC08,SH06,LR77,CWB90}.
Especially in the fields of molecular systems, the relation between avoided crossing and onset of chaos was investigated~\cite{AS10,AB98}.
For convenience of distinction, the avoided crossings in conservative or closed systems are
called avoided level crossings (ALCs) whereas the avoided resonance crossings (ARCs) are their extension to open or dissipative systems \cite{W06}.
For a strong coupling, both cases show similar behaviors but may have different physical origins for avoided crossings.
In particular, the ALCs in closed quantum billiards are due to the symmetry breaking in a block diagonal matrix~\cite{S99,H01}
while the ARCs can be due to the openness effects \cite{R09,D00,PKJ16,YY15}.
Especially, integrable systems like an ellipse or a rectangle can be an interesting platform for studying the ARCs since the openness effects are the sole source \cite{R09,PKJ16} of avoided crossings.
The avoided crossings in quantum billiards have been investigated related to exceptional points~\cite{HS90,SK16,LY09,LR08}, unidirectional emission \cite{RLK09}, dynamical tunneling \cite{SGLX14}, high quality factors \cite{SG13}, ray dynamics \cite{LY009}, and so on.
However, the Shannon entropy, despite its utility, has not been applied to the avoided crossings in quantum billiards to the best of our knowledge.

In this paper, we investigated the relation between the Shannon entropy and the avoided crossings under the strong coupling in dielectric microcavities.
We then found that the Shannon entropy increases due to coherent superposition of wavefunctions as the center of avoided crossing is approached. This result is opposite to the previous one obtained for electrons in atomic systems~\cite{GD033,HC15},
where the Shannon entropy decreases due to electron ionization as we move close to the center of avoided crossing.
In addition, we compared the openness effects and the chaotic effects on the Shannon entropy. For this, we adopted an elliptic billiard as an integrable system for manifesting the openness effects~\cite{PKJ16} and a quadrupole billiard~\cite{CC96} as a non-integrable system for manifesting chaotic effects, respectively.

This paper is organized as follows. In Sec.~\ref{com}, we compare an open quantum system and a closed chaotic quantum system. In Sec.~\ref{ellipse}, we study the Shannon entropy for closed and open elliptic billiards. The Shannon entropy for a closed quadrupole billiard is presented in Sec.~\ref{quad}.  Maximal entropy states and effects of self-energy Lamb shift to Shannon entropy is discussed In Sec.~\ref{max}. Finally, we summarize our work in Sec.~\ref{conclusion}.

\section{Comparison between an open quantum system and a closed chaotic quantum system} \label{com}

The avoided crossing takes place when the off-diagonal terms of Hamiltonian become prominent. These off-diagonal terms arise from various sources depending on the properties of each system. First, let us briefly recapitulate
the avoided crossings due to the openness effects. These openness effects are well described by non-Hermitian Hamiltonian, first developed in nuclear physics~\cite{F58} and then applied to other areas such as atomic physics~\cite{M11}, microwave cavities~\cite{PR00}, solid state physics~\cite{CK09}, dielectric microcavities~\cite{PKJ16,PKJ016}, and so on.

The non-Hermitian Hamiltonian for an open quantum system can be obtained by introducing the Feshbach projective operator $\pi_{S}$ and $\pi_{B}$ with $\pi_{S}\pi_{B}=\pi_{B}\pi_{S}=0$ and $\pi_{S}+\pi_{S}=I_{T}$ . Here, $\pi_{S}$ is an projective operator onto a closed quantum system and $\pi_{B}$ is an projective operator onto a bath, respectively. The operator $I_{T}$ is an identity operator for the total system-bath space. With these operators and Hamiltonian for the total system-bath space $H_{T}$, we define useful operators such as $H_{S}=\pi_{S}H_{T}\pi_{S}$, $H_{B}=\pi_{B}H_{T}\pi_{B}$, $V_{SB}=\pi_{S}H_{T}\pi_{B}$
and $V_{BS}=\pi_{B}H_{T}\pi_{S}$, where $H_{S}$ ($H_{B}$) is a Hamiltonian of the closed quantum system (bath). The $V_{SB}$ denotes an interaction from the bath to the closed quantum system and $V_{BS}$ vise versa~\cite{D00,R09}. The non-Hermitian Hamiltonian is then defined as
\begin{align}
H_\tn{eff}&=H_{S}+V_{SB}G_B^{+}V_{BS},
\end{align}
with an out-going Green function $G_B^{+}$ in a bath. The Green function is defined as $G_B^{+}\equiv(\mu^{+}-H_{B})^{-1}$, where $\mu^+$ is an eigenvalue of $H_{B}$ with a small positive imaginary number added for out-going states: $\mu^{+}\equiv\mu+i\eta$ and $\lim_{\eta\rightarrow0^{+}}$.
If $H_{S}$ with eigenvalue $\epsilon_{j}$ for $j$th eigenstates $\ket{\epsilon_{j}}$ describes an integrable system (no internal interaction), only the second term, $V_{SB}G_B^{+}V_{BS}$, can lead to avoided crossings. In these cases, the out-going Green functions plays a crucial role in the system-bath interaction. That is, the $G_B^{+}$ can route the state not only to the same state but also to a different state $\ket{\epsilon_{j}}\neq\ket{\epsilon_{k}}$, resulting in the collective Lamb shift, otherwise, it is the self-energy Lamb shift. Thus, this interaction via the bath (the collective Lamb shift) can induce an avoided crossing as well as coherent superposition of wave functions~\cite{I13}. On the other hand, the interaction with the bath itself is just giving rise to mode decay~\cite{PKJ16}, not the avoided crossings.

\begin{figure*}
\centering
\includegraphics[width=18.0cm]{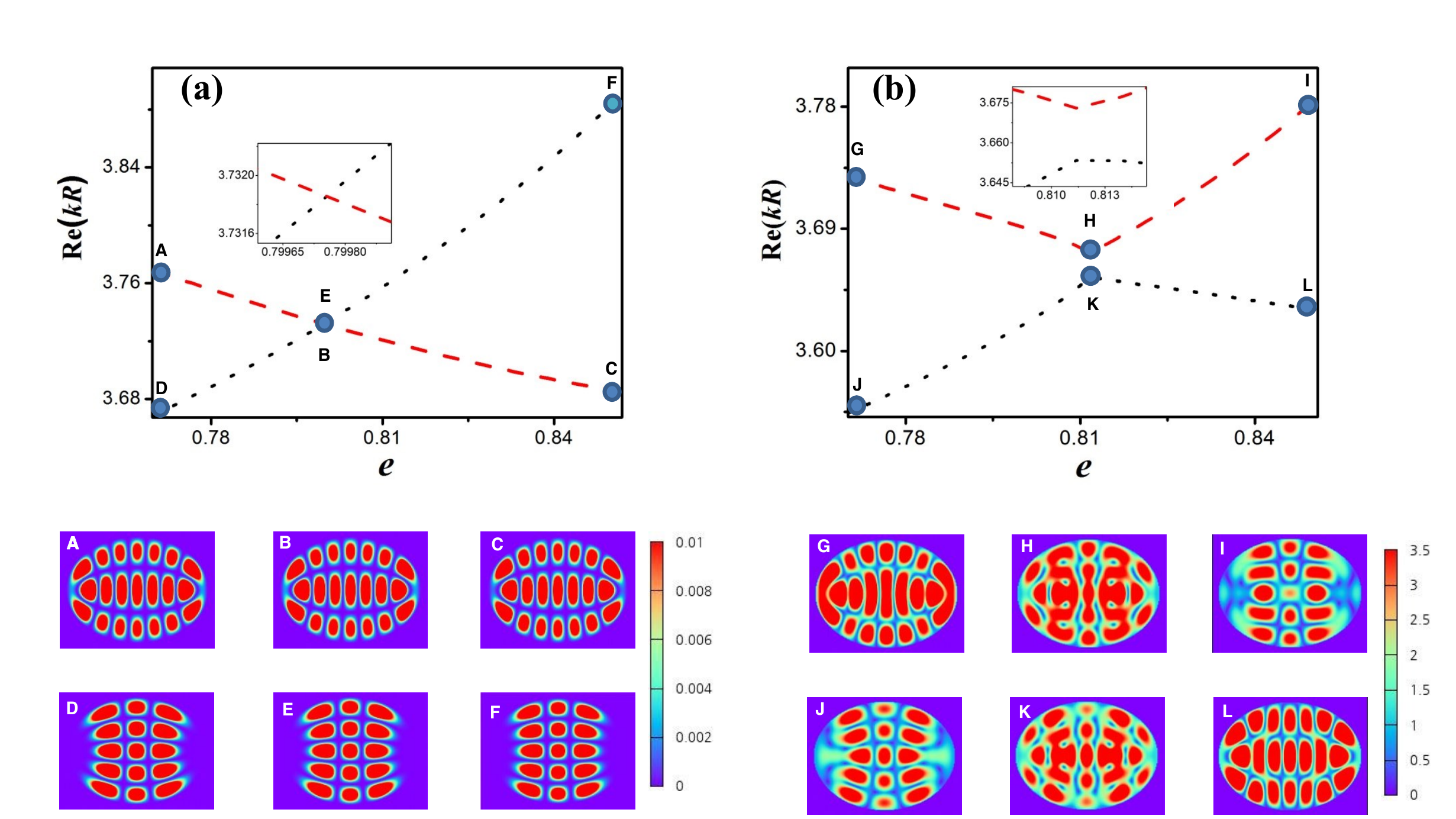}
\caption {(Color online) (a) The real part $kR$ of eigenvalues with their magnified view (inset) near the mode crossing and the intensities of some representative eigenfunctions for a closed elliptic microcavity with the eccentricity $e$. The eigenvalues show a mode crossing near $e\simeq0.805$ and their eigenfunctions are almost unchanged across the mode crossing.  (b) The real part $kR$ of eigenvalues with their magnified view (inset) near the center of avoided crossing and the intensities of some represenative eigenfunctions for an open elliptic microcavity as a function of the eccentricity. The avoided crossing takes place near $e\simeq0.81$ and the eigenfunctions for G and J are mixed at H and K and then exchanged at I and L.
}
\label{Figure-1}
\end{figure*}

Next, let us recapitulate the Hamiltonian properties for a closed chaotic system.
The energy repulsions in quantum chaos are well explained by the random matrix theory~\cite{S99,H01,BG84,GK11}.
When there are $n$-quantum numbers $(O^{1},O,^{2}...,O^{n})$ in a system with $n$-degrees of freedom,
the system is integrable. If not, it is chaotic. This property is fundamentally related to symmetries in the system.
That is, if we can find $(n-1)$ observables which commute with Hamiltonian $H$, {\em i.e.}, $[H,\mathcal{O}^{i}]=0$, giving a complete basis set $\ket {O^{1},...,O^{n}}$, then the Hamiltonian $H$ is fully diagonal with respect to this complete basis set.
If not, the Hamiltonian is block diagonal, having off-diagonal terms. When the symmetries are broken, the size of the block diagonal matrix increases with the number of blocks decreasing.

\section{Comparison between Shannon entropies in closed and open elliptic billiards} \label{ellipse}

Let us consider a closed elliptic microcavity with a major axis $a$ and a minor axis $b$. We are interested in its eigenvalues $\lambda_k$ and eigenfunctions $\psi^{\rm ce}(x)$:
\begin{align}
H^{\rm ce}\psi^{\rm ce}_{k}(x)=\lambda_{k}\psi^{\rm ce}_{k}(x)
\end{align}
\noindent
where the superscript `ce' stands for a closed elliptic billiard with Dirichlet condition $\psi(\textbf{r}=R)=0$ along the boundary. Here $\textbf{r}$ is a position vector and $R$ represents the boundary of the billiard.
We consider transverse-magnetic(TM) modes only with $\psi(\textbf{r})$ corresponding to the electric field.
The eigenvalues are calculated by using the boundary element method (BEM) \cite{W03} with a scanning parameter $\chi$ for $a=1+\chi$ and $b=\frac{1}{1+\chi}$ (constant area).
In Fig.~\ref{Figure-1}(a), the real part $kR$ of eigenvalues are plotted as the eccentricity $e\equiv\sqrt{1-(\frac{b}{a})^2}$ is varied from 0.77 to 0.85 with an interval $\Delta\chi=10^{-5}$. Intensities of some of the corresponding eigenfunctions are also shown.
A mode crossing is observed near $e\simeq0.805$, and their eigenfunctions are almost unchanged across the mode crossing.
This behavior is well known and expected from the random matrix theory. The closed elliptic billiard is an integrable system, and thus it can not lead to avoided crossings, resulting in Poisson distributions~\cite{S99,H01,BG84,GK11}.
However, it was reported that the closed elliptic billiard may lead to Demkov-type interaction in some cases~\cite{kk17}. The Demkov-type avoided crossing occurs over a broad range between two eigenfunctions giving rise to a new pair of eigenfunctions localized on periodic orbits. In contrast, the usual Landau-Zener-type avoided crossing occurs over a short range between two eigenfunctions with exchange of their characteristics.

Next, let us consider an open elliptic cavity.
%
The eigenvalues and eigenfunctions satisfy the time-independent Shr\"{o}dinger equation
\begin{align}
H^{\rm oe}\psi^{\rm oe}_{k}(x)=z_{k}\psi^{\rm oe}_{k}(x)
\end{align}
where the superscript `oe' indicates an open elliptic billiard with boundary conditions $\psi_{in}(\textbf{r}=R)=\psi_{out}(\textbf{r}=R)$ and $\partial_{n}\psi_{in}(\textbf{r}=R)=\partial_{n}\psi_{out}(\textbf{r}=R)$ for TM mode.
For fair comparison with the closed elliptic cavity, we also consider TM modes for the open elliptic cavity.
%
The Hamiltonian $H^{\rm oe}$ can be expressed as \cite{D00,R09}
\begin{align}
H^{\rm oe}= H^{\rm ce}+ V_{SB}G_{B}^{+}V_{BS}
\end{align}
similar to Eq.~(1), exhibiting non-Hermitian properties such as $H^{\rm oe}\neq (H^{\rm oe})^{\dagger}$.

In Fig.~\ref{Figure-1}(b), the real part $kR$ of its eigenvalues $z_k$ and the intensities of some of the eigenfunctions $\psi^{\rm oe}_k(x)$ are plotted as the eccentricity is varied.
An avoided crossing takes place near $e\cong0.81$.
The eigenfunctions corresponding to G and J are mixed at the center of the avoided crossing (H and K), and then exchanged at I and L.
Moreover, the probability distributions (intensity plots) of the mixed eigenfunctions at H and K show more uniform patterns than the unmixed ones at G, J, I and L.
We will use the Shannon entropy to quantify the degree of uniformity.

\begin{figure}
\centering
\includegraphics[width=9.0cm]{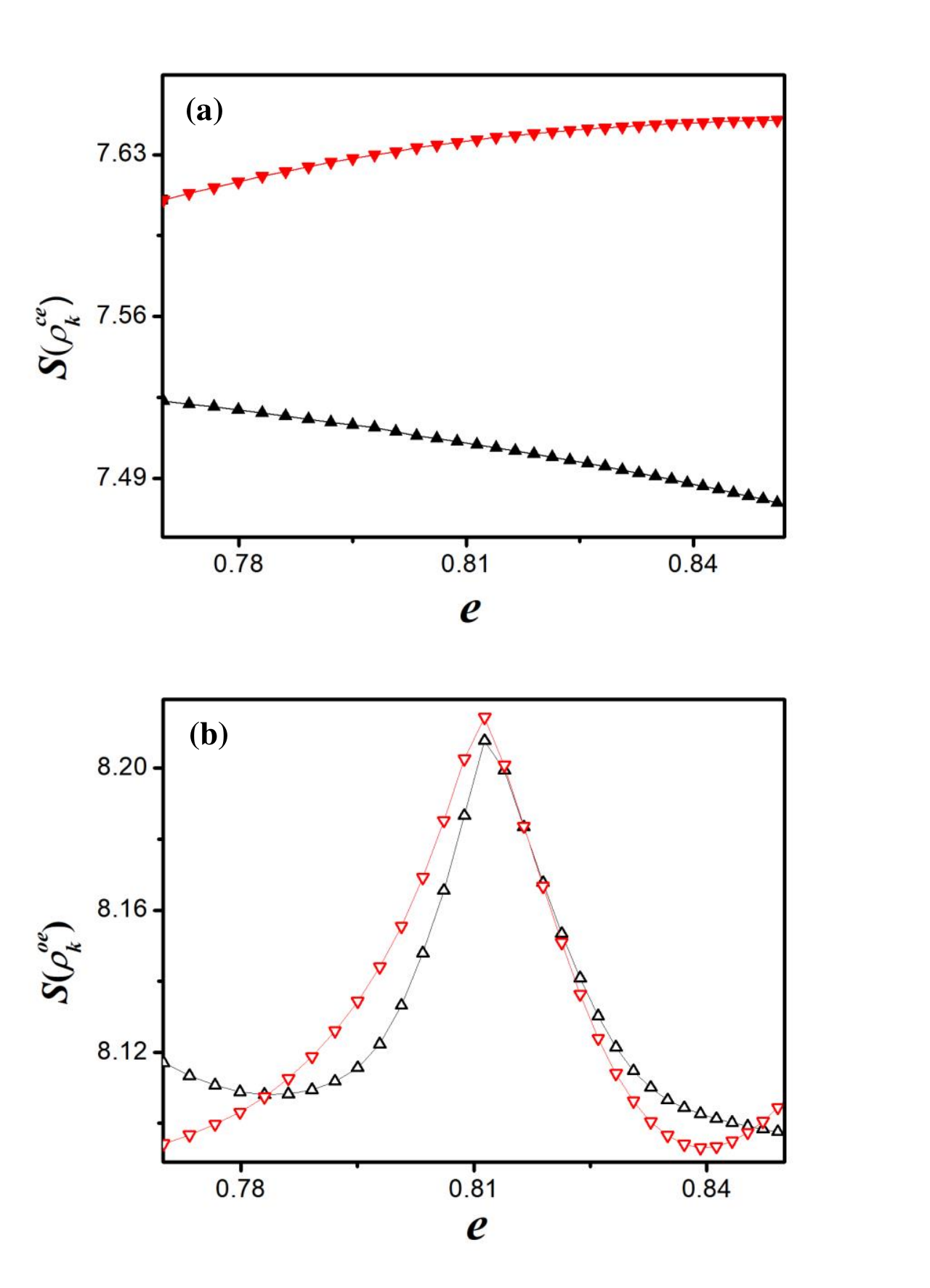}
\caption {(Color online) The Shannon entropy of probability density for elliptic quantum billiards as a function of the eccentricity. (a) The Shannon entropy for Fig.~1(a). The filled red-inverted(black-noninverted) triangles represent the Shannon entropy for the red-dashed(black-dotted) A-B-C(D-E-F) trajectory of the eigenvalues in Fig.~\ref{Figure-1}(a).
(b) The Shannon entropy for Fig.~1(b). The unfilled red-inverted(black-noninverted) triangles represent the Shannon entropy for the red-dashed(black-dotted) G-H-I(J-K-L) trajectory of the eigenvalues in Fig.~1(b).
}
\label{Figure-2}
\end{figure}

For a given discrete probability distribution $\rho(x_i)$ and $N$ number of states, the Shannon entropy is defined as
\begin{align}
S\big( \rho(x_i) \big) \equiv -\sum_{i=1}^{N}\rho(x_{i})\log\rho(x_{i}),
\end{align}
where $\sum_{i=1}^{N}\rho(x_{i})=1$.
We apply this definition to the probability density for a closed elliptic billiard
$S\big(\rho^{\rm ce}_{k}(x_i)\big)$ with a normalization condition $\sum_{i=1}^{N}\rho^{\rm ce}_{k}(x_{i})=\sum_{i=1}^{N} |\psi^{\rm ce}_{k}(x)|^2=1$
for $N$ number of states or mesh points.

The Shannon entropy of probability density is numerically calculated for a closed elliptic billiard as a function of the eccentricity and the result is shown in Fig.~\ref{Figure-2}(a).
The filled red-inverted(black-noninverted) triangles represent the Shannon entropy for a red-dashed(black-dotted) A-B-C(D-E-F) trajectory of the eigenvalues in Fig.~\ref{Figure-1}(a).
We notice that the A-B-C trajectory has larger values of Shannon entropy than the D-E-F trajectory, but both of them change little as the eccentricity is varied.
This result is no surprise since both probability densities remain almost unchanged as the eccentricity is varied and the probability densities of the eigenfunctions on the D-E-F trajectory exhibit a smaller number of antinodes than those on the A-B-C trajectory.

Likewise, we can have the Shannon entropy of probability density for an open elliptic billiard $S(\rho^{\rm oe}_{k}(x_i))$
with normalization condition $\sum_{i=1}^{N}\rho^{\rm oe}_{k}(x_{i})=\sum_{i=1}^{N}|\psi^{\rm oe}_{k}(x)|^2=1$
for $N$ number of states or mesh points.
For the open elliptic billiard considered in Fig.~\ref{Figure-1}(b), the Shannon entropy is numerically calculated and the result is shown in Fig.~\ref{Figure-2}(b) as a function of the eccentricity.
The unfilled red-inverted(black-noninverted) triangles represent the Shannon entropy for a red-dashed(black-dotted) G-H-I(J-K-L) trajectory of the eigenvalues in Fig.~\ref{Figure-1}(b).
It is noted that the behaviors of Shannon entropy of probability density for the open elliptic billiard is quite different from that for the closed one.
The Shannon entropy values of two eigenstates in the open elliptic billiard are maximized at the
center of avoided crossing. Moreover, they are exchanged across the avoided crossing. The Shannon entropy (unfilled black-uninverted triangles) associated with the eigenstate on the J-K-L trajectory is larger than that (unfilled red-inverted triangles) on the G-H-I trajectory for small eccentricity $e\sim 0.77$.
However, after going through the center of the avoided crossing, the Shannon entropy denoted by the unfilled black-uninverted triangles becomes smaller than that represented by the unfilled red-inverted triangles for large eccentricity $e\sim 0.85$.
This exchange of Shannon entropy is consistent with the intensity pattern exchange shown in Fig.~\ref{Figure-1}(b).

It should be noted that the collective Lamb shift becomes dominant over the self energy for large eccentricity in an open elliptic billiard as shown in Fig.~\ref{Figure-1}(b), resulting in an avoided crossing.
At the avoided crossing, two interacting eigenfunctions are mixed together, resulting in
more uniform probability distributions and consequently increased Shannon entropy as shown in Fig.~\ref{Figure-2}(b).

\section{Shannon entropy in closed quadrupole billiards} \label{quad}

\begin{figure}
\centering
\includegraphics[width=9.0cm]{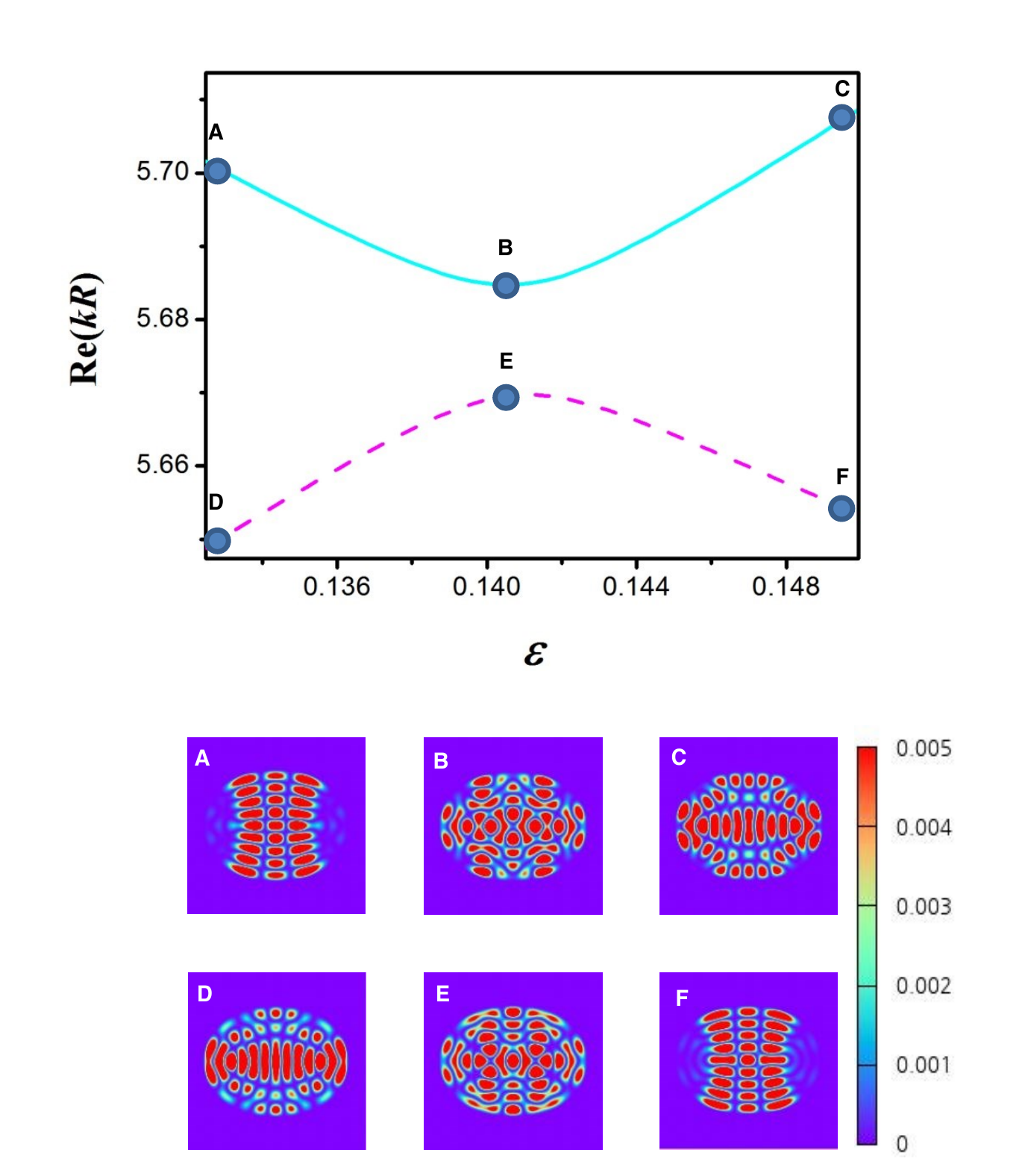}
\caption{(Color online) The trajectories of eigenvalues and the intensities of some representative eigenfunctions in a closed quadrupole billiard. The upper panel shows the trajectories of eigenvalues with the deformation parameter $
\varepsilon$. The avoided crossing takes place near at $\varepsilon\simeq0.141$. The lower panel shows that eigenfunctions for A and D are mixed together B and E, and then undergo mode exchange at C and F.
}
\label{Figure-3}
\end{figure}

In this section, we consider a closed chaotic billiard.
In particular, we consider a quadrupole described by
$r(\phi)=1+\varepsilon \cos(2\phi)$ with a deformation parameter $\varepsilon$.
The eigenvalues $\lambda_{k}$ and eigenfunctions $\psi^{\rm cq}_{k}(x)$ satisfying
\begin{align}
H^{\rm cq}\psi^{\rm cq}_{k}(x)=\lambda_{k}\psi^{\rm cq}_{k}(x),
\end{align}
are numerically calculated by using BEM~\cite{W03} with an interval of $\Delta\varepsilon=10^{-5}$.
The superscript `cq' stands for `closed quadrupole'. We consider TM modes with Dirichlet boundary condition.
The resulting trajectories of eigenvalues and the intensities of some of the eigenfunctions are shown in Fig.~\ref{Figure-3} as $\varepsilon$ is varied from 0.134 to 0.149. It is seen that an avoided crossing takes place near $\varepsilon\cong0.141$ as in an open elliptic billiard even though their origins are completely different from each other as already discussed in Sec.~\ref{com}.
Two distinct eigenfunctions at A and D are mixed together at B and E and then undergo a mode exchange
at C and F.
This type of avoided crossings is due to the symmetry breaking in a block diagonal Hamiltonian for a closed quantum chaotic system.
It is seen that the probability distributions of the mixed eigenfunctions at B and E reveal more uniform patterns than the unmixed ones at A, D, C, and F.
Note that $\psi^{\rm cq}_{k}(x)$ describing a chaotic system cannot form a complete basis set \cite{S99,H01}.

\begin{figure}
\centering
\includegraphics[width=9.0cm]{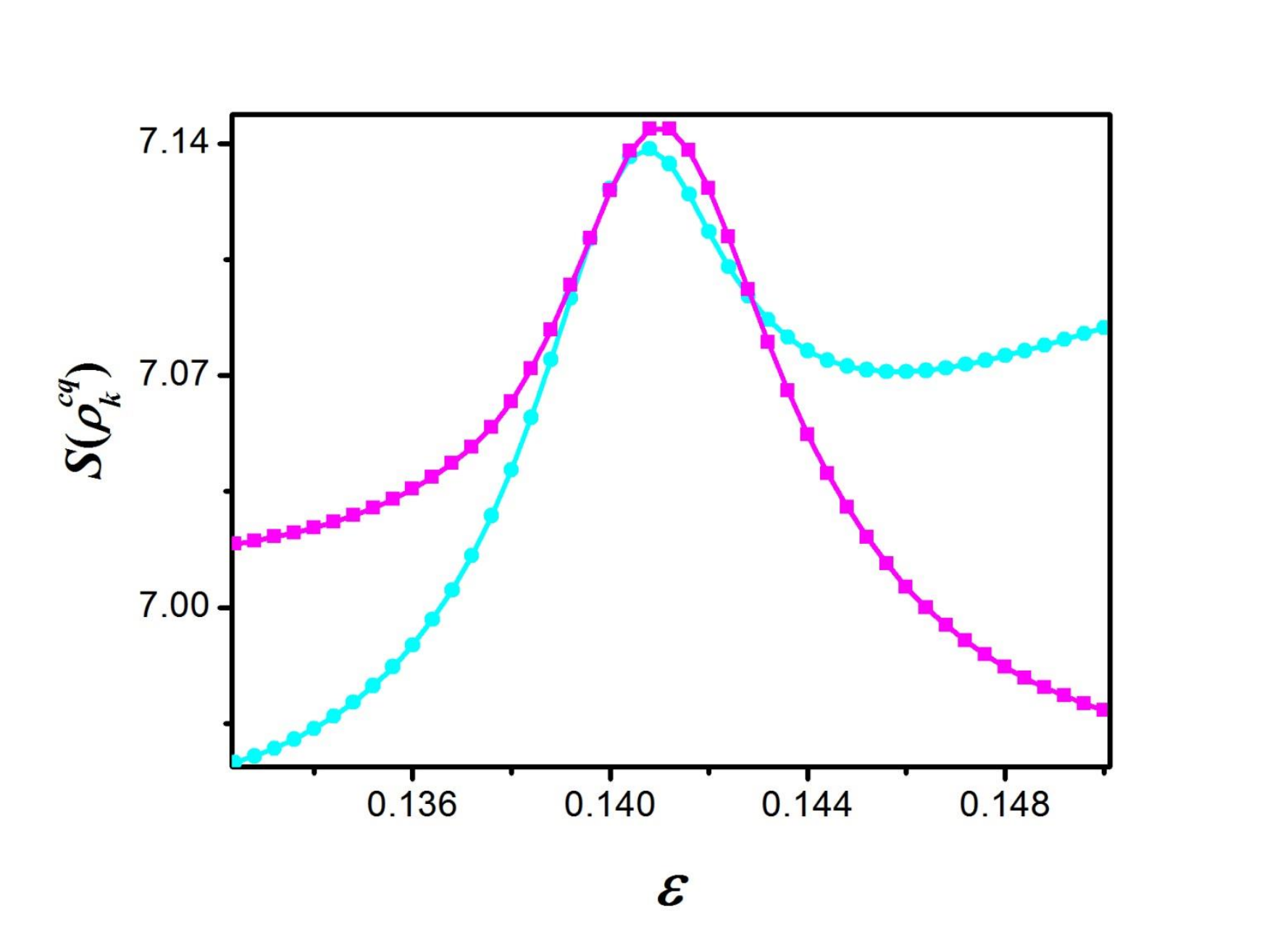}
\caption{(Color online) The Shannon entropy in a quadrupole quantum billiard. The cyan circles(magenta squares) represent the Shannon entropy for the cyan solid(magenta dashed) eigenvalue trajectory in Fig.~\ref{Figure-3}. The Shannon entropy are maximized near the center of avoided crossing ($\varepsilon\simeq0.141$) and they are exchanged across the avoided crossing.
}
\label{Figure-4}
\end{figure}

Figure 4 shows the Shannon entropy of probability density for the quadrupole quantum billiard $S\big(\rho^{\rm cq}_{k}(x_i)\big)$
with normalization condition $\sum_{i=1}^{N}\rho^{\rm cq}_{k}(x_{i})=\sum_{i=1}^{N}|\psi^{\rm cq}_{k}(x)|^2=1$ for $N$ number of states or mesh points.
The cyan circles(magenta squares) represent the Shannon entropy for the cyan solid(magenta dashed) eigenvalue trajectory in Fig.~\ref{Figure-3}.
The Shannon entropies are maximized near the center of avoided crossing ($\varepsilon\simeq0.141$) and they are exchanged across the avoided crossing, similar to the Shannon entropies in the open elliptic billiard in Fig.~\ref{Figure-2}(b).

\section{Maximal entropy state and effect of self energy}\label{max}

\begin{figure}
\centering
\includegraphics[width=9.0cm]{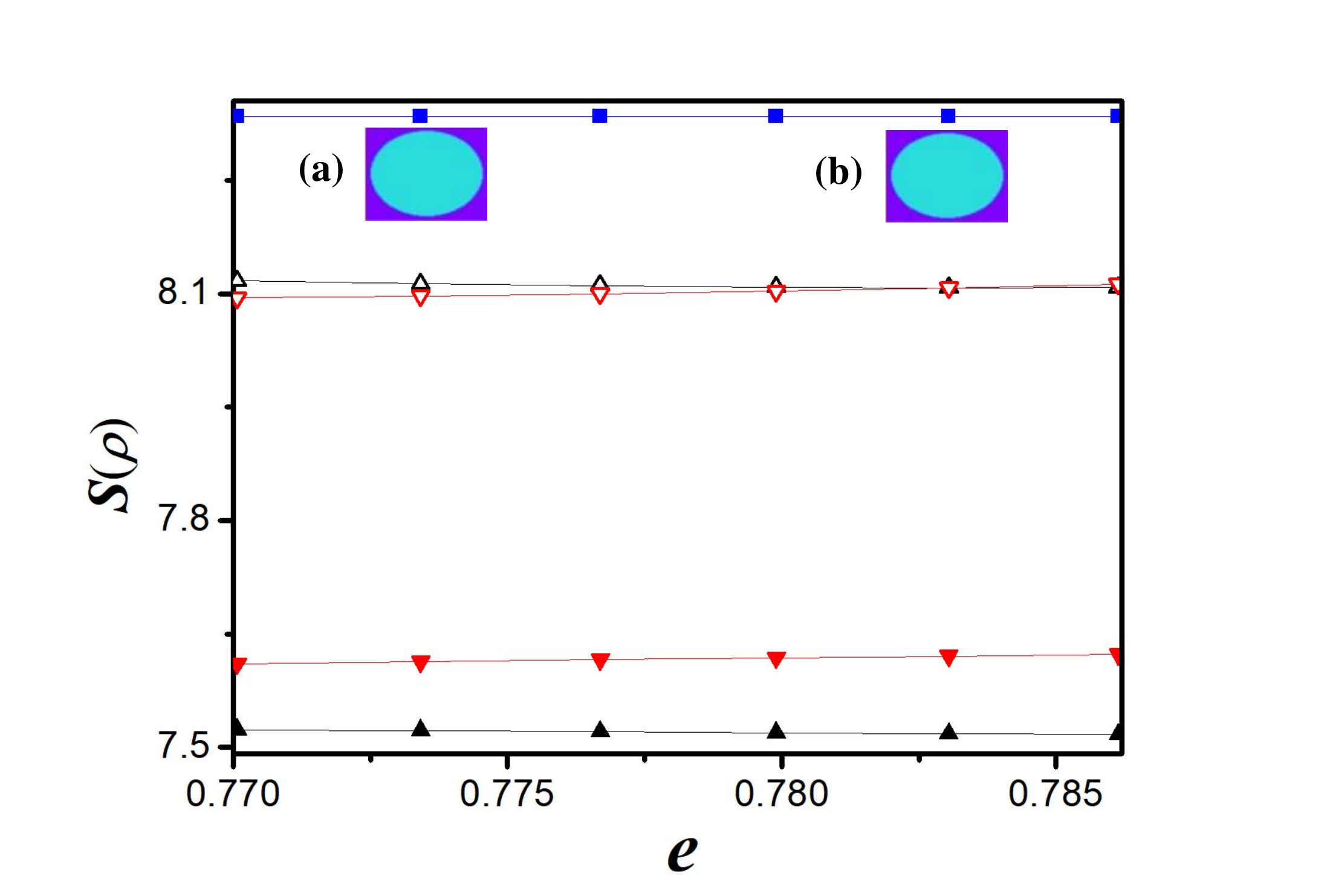}
\caption{(Color online) The Shannon entropy in Fig.~2(a) and Fig.~2(b) are shown in a different range from $e\cong 0.770$ to $e\cong0.785$ well before the avoided crossing. A maximum entropy is also presented by blue squares. Unfilled triangles are the Shannon entropy for the open elliptic billiard whereas filled triangles are that of the closed elliptic billiard. The insets (a) and (b) show the maximal entropy states (uniform distribution states) at $e\approx0.773$ and $e\approx0.783$, respectively.
}
\label{Figure-4}
\end{figure}
Even though Fig.~2(b) and Fig.~4 shows similar behaviors, that is, the Shannon entropy is maximized as the center of avoided crossing is approached and they are exchanged across the avoided crossing in both cases, the detailed behaviors are slightly different from each other. The relative difference between two curves away from the center of avoided crossing in Fig.~2(b) is small compared to Fig.~4 and the behaviors of two curves are more complicated (curves cross each tree times) than Fig.~4. These slightly different behaviors can be explained when we consider a maximal entropy state and the self energy.

A uniform probability distribution $q(x)=\frac{1}{N}$ for $N$-number of states is a special one since it gives rise to a maximal entropy, {\em i.e.}, $S(\rho_{\tn{max}})=\log{N}$. This maximal entropy can be obtained in a quantum billiard
by identifying the number of states $N$ with mesh points for numerical calculation of system.
Blue squares in Fig.~5 represent  a maximal entropy for an elliptic cavity and it has a constant value as $S(\rho_\tn{{max}})\simeq8.381$ with $N=4166$ for the number of states. Note that the states for the maximal entropy are not eigenstates of the Hamiltonian:
we artificially impose uniform intensities in billiard systems.
Obviously, if the mesh point $N$ is fixed as a constant value,
the maximal entropy is also fixed regardless of the deformation parameter. The insets (a) and (b) show the maximal entropy states (uniform distribution states) at $e\approx0.773$ and $e\approx0.783$, respectively.

To examine the self energy (mode decay) effects to Shannon entropy, we vary the eccentricity from $e\approx0.770$  to $e\approx0.785$, well before the avoided crossing, as shown in Fig.~5.
The wavefunctions at boundaries are zero in closed billiards (Fig.~1a) whereas those are non-zero in open billiards (Fig.~1b).

This fact indicates that the self-energy Lamb shift makes the wavefunctions themselves dispersed, making them closer to the maximal entropy state.
As a result, the absolute values of Shannon entropy for wavefunctions in an open billiard become larger than those in a corresponding closed billiard.
At the same time, the relative difference between the two curves of Shannon entropy in the open billiard becomes smaller than that in the closed billiard. This fact implies that not only the collective Lamb shift but also the self-energy can induce a change in the Shannon entropy.
By these reasons, the detailed behaviors of Fig.~2(a) for an open billiard are different from those of Fig.~4 for a closed billiard.
Since we focus on the relation between the avoided crossing and Shannon entropy in this paper, the effect of the self energy on Shannon entropy will be discussed elsewhere.

\section{Conclusion} \label{conclusion}

We investigated the relation between the Shannon entropy and avoided crossings under strong coupling in dielectric microcavities.
Before our works, the relation between the Shannon entropy and avoided crossing was investigated in atomic physics and their result was opposite to ours, {\em i.e.}, the Shannon entropy for electron \emph{decreases} due to electron ionization as we move close to the center of avoided crossing. On the contrary, the Shannon entropy of probability density for dielectric microcavities (quantum billiards) \emph{increases} due to the coherent superposition of wavefunctions as the center of avoided crossing is approached, but both cases show exchanges of Shannon entropy as well as mode exchanges. The Shannon entropy of probability density for a closed elliptic billiard changes little with the eccentricity while the Shannon entropy of probability density for an open elliptic billiard is maximized at the center of avoided crossing.
This maximization and exchange of Shannon entropy in an open elliptic billiard comes from the collective Lamb shift, which is an energy level shift due to the interaction of energy levels with each other via the bath and can also induce a avoided crossing and coherent superposition of wavefunctions.
In a closed quadrupole billiard, the Shannon entropy is also maximized as the center of avoided crossing is approached with both exchange of Shannon entropies as well as mode patterns.
This maximization and exchange of Shannon entropy in a closed quadrupole billiard comes from the nonlinear dynamical effects in a chaotic system.
Irrespective of origin of the avoided crossings, the open elliptic cavity and the closed quadrupole cavity show similar behaviors to the Shannon entropy. That is, the collective Lamb shift of open quantum systems and the symmetry breaking of chaotic quantum systems give equivalent effects to the Shannon entropy.

\section{acknowledgement}
We are grateful to Myung-Woon Kim, Sunghwan Rim for comments. This work was supported by
the Korea Research Foundation (Grant No. 2016R1D1A109918326). K.J. acknowledges financial support by the National Research Foundation of Korea (NRF) through a grant funded by the Korean government (Ministry of Science and ICT) (NRF-2017R1E1A1A03070510 and NRF-2017R1A5A1015626).

\end{document}